# GENIE OBSERVATIONS OF SMALL SCALE ASTROPHYSICAL PROCESSES IN STAR FORMING REGIONS AND QUASARS

Glenn J. White [(1)], Stephen B. G. Serjeant[(1)]

[(1)]School for Physical Sciences,
University of Kent,
Canterbury CT2 7NR, England
Email:g.j.white@ukc.ac.uk

## INTRODUCTION

The VLTI/GENIE configuration will operate using at least 4 of the VLTI telescopes (and possibly with one or more of the AT telescopes in the future if adaptive optics become available on them). GENIE effectively can be thought of as a 'smart' coronagraph, enabling high dynamic range imaging to be achieved at moderate spatial resolution, with high rejection of the emission of a central bright point source. However, but this bright source rejection may only provide a rather moderate image quality (due to the few baselines and transfer function on the sky). Operated in this way, only limited image reconstruction is possible since classical radio and millimeter wavelength interferometry techniques are not directly applicable to the outputs of optical interferometers because the absolute phases are generally not measured. However, measurements of visibility and closure phase could lead to situations where image reconstruction becomes possible. This paper addresses the issue of whether there areas outside of the exoplanet search where it might be able to make a useful impact on astronomy.

## INTRODUCTION TO NULLING INTERFEROMETRY

The basic concept of nulling interferometry is conceptually simple: the light incident on a pair of telescopes is combined so that at zero optical path difference between the incident beams, the two electric vectors are exactly 180 out of phase - allowing near-perfect starlight subtraction. A deep destructive fringe is to be placed across the star – or central bright object – but emission from sources located near constructive fringe maxima can be transmitted through the system, and so even though the star is nulled to a deep level, features lying slightly displaced from the central position are not attenuated greatly. The configuration of a single GENIE interferometer pair is shown in Fig. 1.

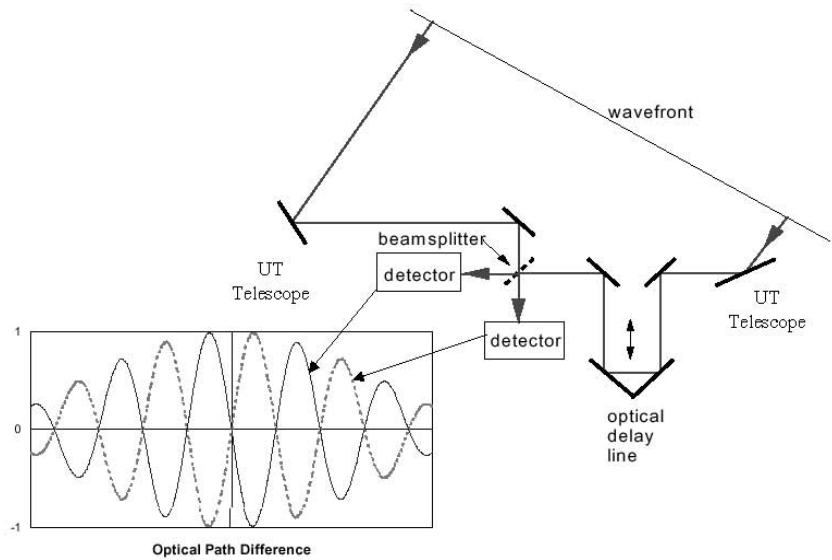

Fig. 1. Outline of the Nulling Interferometry configuration using two of the ESO UT telescopes

The transmission of a Bracewell interferometer can be readily calculated. Assuming that a single interferometer pair consists of two telescopes of aperture *D* separated by a baseline *B*, the transmission *T* can be expressed as a function of angular distance *θ* and azimuth *φ* with respect to the interferometer baseline projected onto the sky using the relationship:

$$T(\theta,\varphi) = \left[2 \cdot \frac{J_1\left(\frac{\pi \cdot \theta \cdot D}{\lambda}\right)}{\left(\frac{\pi \cdot \theta \cdot D}{\lambda}\right)}\right]^2 \cdot \sin^2\left(\frac{\pi \cdot B}{\lambda} \cdot \theta \cdot \cos(\varphi)\right) \tag{1}$$

Assuming that the orientation of the target structure is known – for example a micro jet emanating from the heart of a quasar, or the root of a jet associated with a molecular outflow around a young star, is known – then it is possible to observe the source with the optimum combination of baseline separation and hour angle coverage to be sensitive to that structure. Fig. 2 shows a classical interferometric simulation of the observation of a gaussian point source, showing the *u-v* tracks on the plane of the sky, and the reconstruction from the sparsely sampled data that can be achieved.

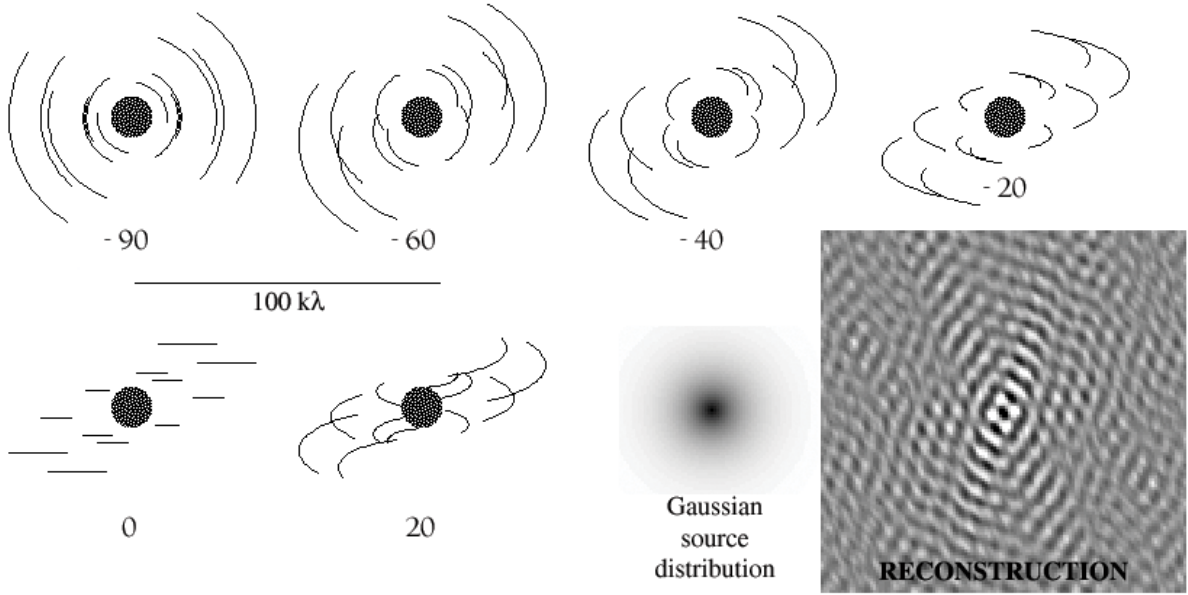

Fig. 2. *UV* coverage of the GENIE array of four UT telescopes, and the classical interferometric reconstruction of a compact gaussian point source

Although the main science goal for the GENIE instrument will be to search the DARWIN target sources, in characterising exozodial dust, and in the search for hot Jupiter's or low luminosity stellar companions. There are several limitations to the use of GENIE for astrophysical imaging. First is the very sparse *uv*-plane coverage shown in Fig. 2, which limits the complexity and dynamic range of fields that can be imaged. This problem of the limited *uv*-plane coverage could be addressed in a number of ways in the design of the interferometer: multi-wavelength baseline synthesis could be used to increase the *uv*-plane coverage of spectrally simple sources; and use of both UT and AT telescopes simultaneously. A second drawback of GENIE is its limited field of view, which is approximately that of the Airy disc of an individual telescope, which is typically ~ 2 arc seconds. This latter limited field of view restriction could be somewhat ameliorated by mosaicing images obtained with multiple pointings and taking advantage of the intrinsically high sensitivity of the instrument. In this paper we note several kinds of observations that could benefit from the characteristics of the GENIE system: a) mapping of proto-planetary disks in nearby star-forming regions – particularly the small scale structure of highly collimated outflow jets and material at the edges of the accretion discs to study their energetics and the role of stellar vs. accretion infall heating; b) With a resolution of 0.1 to 1 pc, the scale of the broad-line region, it will be possible to study the bright line regions and starburst and AGN energy sources in infrared luminous galaxies - compared to more classic optical interferometric observations, GENIE will be able to search for faint structures sited very close to a bright point-like central sources that would otherwise completely dominate the ability to see the expected faint structures; c) observations of lensing in the vicinity of quasars.

**MICROLENSING**

Micro-lensing is an exciting technique that allows us to study the nature of dark matter in our Galaxy. Existing micro-lensing surveys are now finding dozens of star-star lensing events in a year. Expanded surveys will yield many more such lensing events. However, existing surveys provide only photometric information about the light curve of the lensed source. The nature of dark matter is one of the outstanding puzzles in modern astrophysics. If a large fraction of dark matter is made of MACHOs, then the arrival of large ground based interferometers such as GENIE would be most useful to existing (and upcoming) microlensing experiments, in tackling the mass distribution of dark matter.

There are now a number of sensitive dedicated microlensing searches that are being set up, or that are currently operational. Apart from area and time coverage, the use of automatic detection algorithms aim to provide rapid indications of microlensing events – raising the possibility that VLTI/GENIE may could be used in a responsive mode to study lensing events. It is well established that when a dark mass, like a brown dwarf or another massive but dark object (a MACHO), passes in front of a background star, the light from the star is gravitationally lensed. The scale of the lensing is not large enough for two or more distinct images to form, as seen in strong lensing events, but the star's light can be significantly magnified/amplified. However, nulling interferometry does offer a possibility to start to probe much closer in towards the bright stellar object that is being occulted – particularly if this were a QSO, and raises the possibility of being able to directly observe microlensed images with GENIE. The classical detection of microlensing has come from photometric observations as illustrated in Fig. 3:

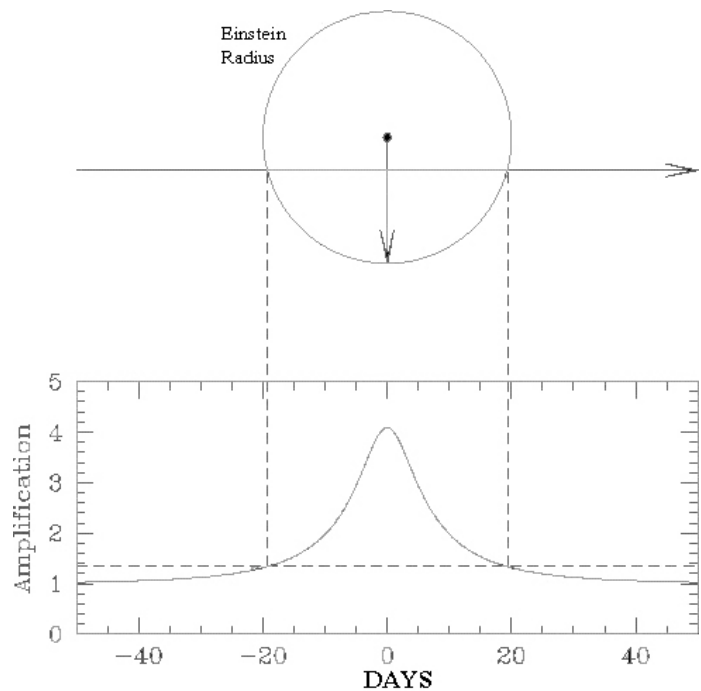

Fig. 3. Variation of the light amplification as a function of impact parameter and Einstein radius

Microlensing refers to the case where the lens is a star. For a star lens the separation between the multiple images of the source is too small to be resolved. Lensing magnifies the source, making it brighter. Therefore as the lens moves in front of the source, the source brightens, and then fades. Since the brightening depends on the mass of the lens, microlensing has successfully been used by several groups (MACHO, EROS, OGLE) to investigate the dark matter content of the Milky Way galaxy. When a mass, like a brown dwarf or another massive but dark object (a MACHO), passes in front of a background star, the light from the star is therefore said to be gravitationally lensed. The scale is not large enough for two or more separate images to form, like in strong lensing events, but the starlight can be significantly magnified/amplified. This amplification can produce a sharp peak in the intensity of light from the background star. If there is a perfect Observer-Lens-Source alignment, the solution to the lens equation describing gravitational lensing around a point mass picks shows spherical symmetry, and a whole ring of solutions are seen. The optics of the system also produces a large amplification of the source signal on this ring. In fact, this ring is a *critical line* of the optical system. The corresponding caustic is a single point on the source plane. The geometric configuration is shown in Fig. 4.

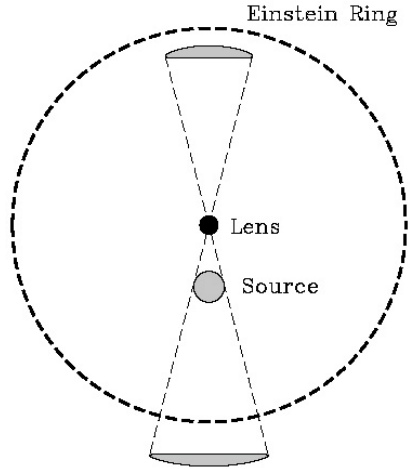

Fig. 4. Geometric configuration of a lensed configuration

A gravitational lens can form multiple images of the source. These are point images for a point source such as a quasar, but a galaxy, being extended, is lensed into a set of arcs, or may appear as a ring. In Fig. 5 we calculate the expected size of the Einstein Ring for a range of dark mass sizes located at different distances.

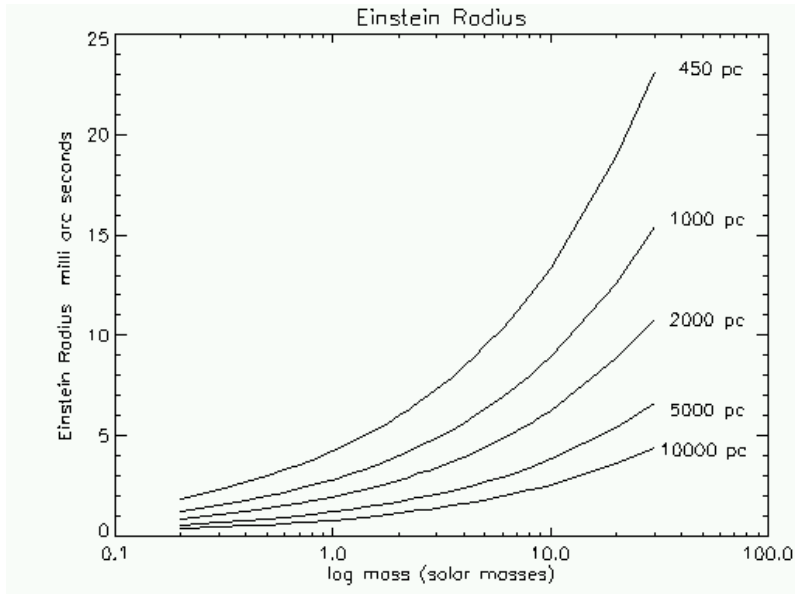

Fig. 5. Einstein radius as a function of distance to and mass of the dark object

Fig. 5 shows that nearby micro-lensed events would be relatively easy to detect with the GENIE resolution – however, the number of events would be so small as to rule this out. This is why most ground-based searches are focusing on the Magellanic Clouds and Galactic Centre as background sources, with lensing being done by sources in the bulge, or along the line of sight. From the figure to the left, we see that the lensed images in this case remain small. For GENIE the 47m baseline in the L band would give a spatial resolution of ~ 15 mas, and a 130m baseline (using one of the AT telescopes) in L ~ 5.4 mas. However, a major limitation would probably be stellar leakage and background subtraction.

**AGN**

The compact cores of bright active galactic nuclei (AGN's) can dominate the luminosity of many galaxies. High spatial resolution infrared imaging with both GENIE could result in very significant progress in our understanding of AGN's enabling obscured star clusters and bright individual stars can be studied in great detail; the distribution and dynamics of interstellar gas disks and flows on scales of a few parsecs or less can be investigated and the environments of massive central black-holes to be studied on scales of less than $10^3$ Schwarzschild radii. Many AGN's are believed to contain black holes, which give rise to a number of energetic phenomena that can only be studied by observing material directly adjacent to the cores. The basic geometry of the central region of an AGN is shown in Fig. 6:

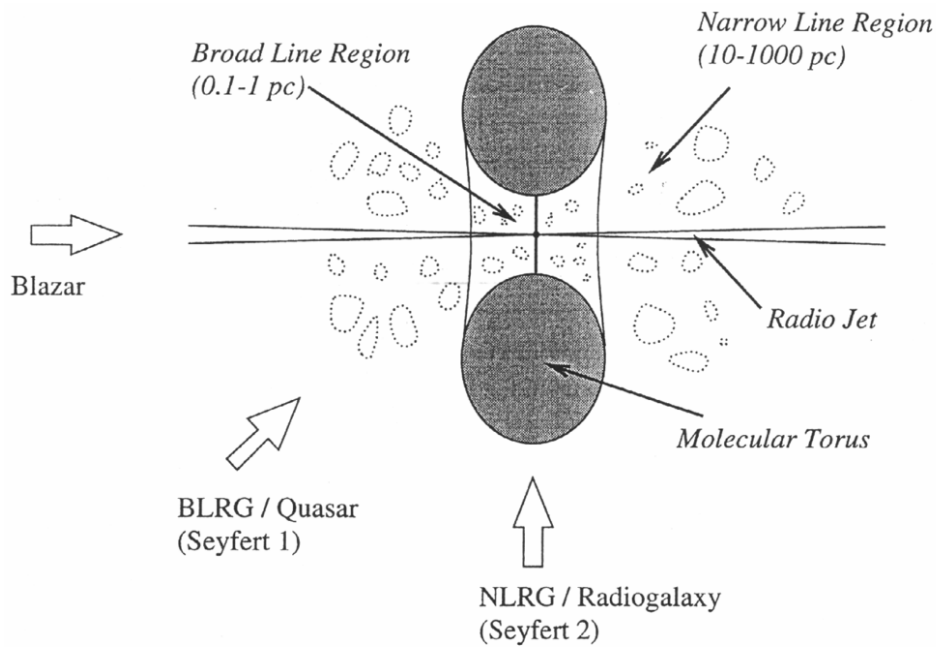

Fig. 6. Basic geometry at the centre of an AGN

The commonly accepted view of AGN is that they are powered by the exchange of gravitational and thermal energy in a compact accretion disk surrounding a central black hole BH. In the unified model of AGN, they accretion disc is surrounded by an optically thick torus composed of dust and molecular and neutral gas that hides the central region from our view, making the direct high energy phenomena associated with the nucleus more difficult to observe. Then, the object does not show the broad lines and is classified as Type II. If the object is viewed from close to the polar axis, a very bright nucleus is observed and the optical spectrum shows the broad lines. It will be classified as Type I. The important issues are whether dust tori in AGN exist at all and whether their physics can be constrained. The orientation of this with respect to our line of sight determines whether we see the object as a Type I ( Seyfert 1 or quasar ) or Type II ( Seyfert 2 or radio galaxy) (see Fig. 6).

As an example, the nearest bright quasar, 3C 273 has a B magnitude <13 , and lies at a redshift z = 0.158 (1" = 2.37 pc and the half-light radius $R_{1/2}$ ~ 5 – 20 Kpc. There are ~ 1500 QSO's with B<18. The inner edges of the circumstellar discs in star forming regions are expected to be ~ 0.5 AU, which in the closest cases corresponds to ~ 4 mas. The inner radii of the dust tori of AGN's are typically ~ 1 pc from the central source, which corresponds in the closest cases to ~ several tens of mas. For nearby AGN, the scale of the BLR ~ 1 light month at z ~ 0.01 ~ 0.2 mas – below that achievable with GENIE (or VLTI). Giant dusty tori surrounding the cores can be readily recognised from spectroscopy and polarimetry. Seyfert 1 Gals / QSO's where the nuclear continuum and BLR are directly visible. Seyfert 2 narrow line radio galaxies where the components are only seen in scattered light. The inner edges ~ 1 pc from the core ~ 20 mas at z = 0.01. NGC1068 has K magnitude ~ 9.3 / inner edge ~ 10 mas. The narrow line emission region extends out to 50 – 100 pc. In the figure below we show predictions of the source sizes of the various components in AGN's as a function of redshift – not that the sizes increase for redshifts > 1.5. Therefore GENIE will be able to observe the narrow line and extended emission line regions in many QSO's, and the BLR and torus in the closest galactic nuclei.

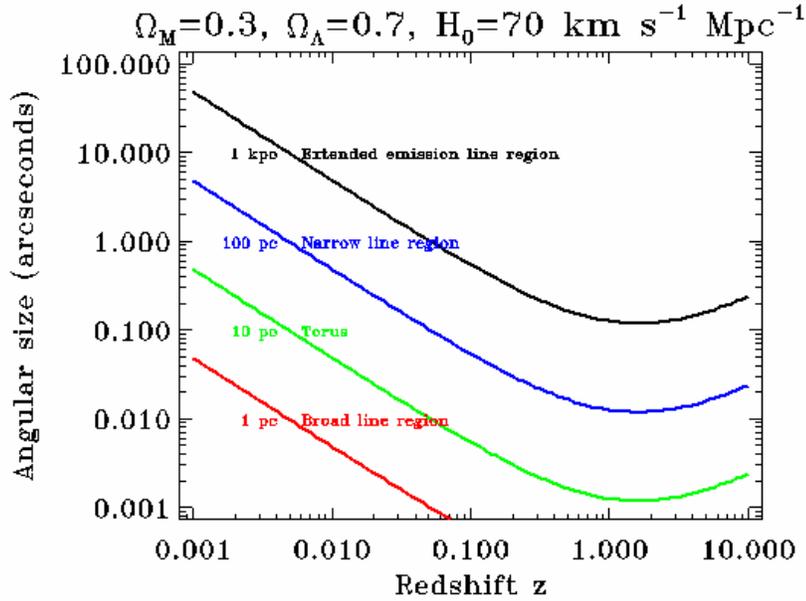

Fig. 7. Sizes of various components of an AGN as a function of redshift, with $H_o = 70$, $\Omega_{matter} = 0.3$ and $\Omega_\lambda = 0.7$

**CONCLUSIONS**

It is clear that the main scientific drivers for GENIE remain on the characterisation of exozodiacal dust, and as a technology demonstrator for the future DARWIN mission. However, we have pointed towards a range of astronomical phenomena: highly collimated jets in star forming region; gravitational microlensing near quasars, and studies of material close to the centres of AGN's that could be benefit from observations with the GENIE interferometer.